\documentclass[10pt,letterpaper]{article}
\usepackage{opex3}
\usepackage{cite}
\usepackage{amsmath}

\begin{document}

\title{Efficient Hotspot Switching in Plasmonic Nanoantennas using
  Phase-shaped Laser Pulses controlled by Neural Networks}

\author{Alberto Comin$^*$ and Achim Hartschuh}

\address{Department of Chemistry and Center for NanoScience (CeNS), LMU
  Munich, 81377, Munich, Germany}

\email{$^*$alberto.comin@cup.uni-muenchen.de}




\begin{abstract*} We present a novel procedure for manipulating the near-field
  of plasmonic nanoantennas using neural network-controlled laser
  pulse-shaping. As model systems we numerically studied the spatial
  distribution of the second harmonic response of L-shaped nanoantennas
  illuminated by broadband laser pulses.  We first show that a trained neural
  network can be used to predict the relative intensity of the second-harmonic
  hotspots of the nanoantenna for a given spectral phase and that it can be
  employed to deterministically switch individual hotspots on and off on
  sub-diffraction length scale by shaping the spectral phase of the laser
  pulse.  We then demonstrate that a neural network trained on a
  $90~\mathrm{nm} \times 150~\mathrm{nm}$ nano-L can in addition efficiently
  predict the hotspot intensities in an antenna with different aspect ratio
  after minimal further training for varying spectral phases. These results
  could lead to novel applications of machine-learning and optical control to
  nanoantennas and nanophotonics components.
\end{abstract*}


\bibliographystyle{osajnl}
\bibliography{coherentControlLs}


\section{Introduction} \label{sec:introduction}

Plasmonic nanoantennas can efficiently focus broadband optical fields to form
nanometer sized ultrafast hotspots~\cite{Novotny2011, Piatkowski2016}. They
could thus become a key component of future nanophotonic devices, combining
high storage density and fast processing of information~\cite{Stockman2011,
  Gu2014, Ashall2013, Stockman2018}.  The ability to deterministically control
the brightness of individual hotspots within a nanoantenna by varying the
input optical fields would add enormous flexibility to antenna schemes. Towards
this goal, several groups have explored the possibility of controlling the
near-field in plasmonic structures by tuning the spectrum, polarization and
spectral phase profile of the input laser
excitation~\cite{Brixner2006,Piredda2010,Stockman2011, Aeschlimann2010,
  Brinks2013, Kojima2016}.

Spectral amplitude shaping could be applied in the simple case in which the
nanoantenna features spectrally distinct plasmonic resonances connected to
hotspots at different positions. For such nanoantennas, it is possible to lit
each hotspot individually by tuning the color of the excitation light, albeit
at the expense of reduced spectral bandwidth and thus temporal resolution. In
more general cases, spatio-temporal control of plasmonic near-fields is
understood to have two main control mechanisms~\cite{Brixner2006}.  Efficient
tuning can be achieved by polarization pulse shaping which exploits the
interference of plasmonic near-field modes with different polarization
responses.  Polarization pulse shaping has wide applicability and was
successfully used to experimentally demonstrate sub-wavelength hotspot
switching on different antenna systems~\cite{Aeschlimann2012,Brixner2006}.
The second control mechanism is based on spectral phase shaping without the
need of polarization control. In this scheme, the spatial control of
non-linear responses is achieved by imprinting a spectral phase profile on the
incident laser pulses which is set to compensate the phase response of the
nanoantenna at a particular position~\cite{Brinks2013,Huang2009,
  Brixner2006a}.

Whereas for simple systems the parameters for the coherent control of the
hotspot position can mostly be derived analytically, the optimum pulse
characteristics for more complex multi-modal systems cannot be
predicted~\cite{Tuchscherer2009}. Although evolutionary algorithms, such as
genetic algorithms (GA), could be employed as versatile optimization tools for
such problem sets, they typically require a large number of experimental iterations
and are thus limited in efficiency~\cite{Baumert1997}.  Furthermore, the
results obtained by GA are hard to generalize and replicate, while the GA must
be re-optimized for each sample and experimental configuration. For example,
when an external perturbation shifts the plasmonic resonance of the
nanoantenna, the GA can only adapt by crossover and random mutation, without
taking into account previously learned information about the
sample~\cite{Meyer-Baese2014}.

We propose a combined use of GA and neural-network (NN) as a more efficient
and deterministic way to control the near-field in plasmonic nanoantennas. NN
have a layered structured which can encode more general information in the
first layers and more specific information in the last
layers~\cite{Goodfellow-et-al-2016}. This is one of the main reasons why NNs
became very popular: It is possible to re-use a pre-trained NN on a different
domain after just minimal training of the final
layers~\cite{Goodfellow-et-al-2016}.

In the following, we will show that a NN consisting of only four fully
connected layers can accurately encode the dependence of the near-field on the
spectral phase of the incident laser pulses.  ~\cite{Brinks2013,Huang2009,
  Brixner2006a}.  Moreover, a NN trained on a specific nanoantenna can also be
used with minimal further training on nanoantennas with different size and
aspect ratio.

As an example of the efficacy of this approach, we apply the GA-NN combination
to achieve second harmonic generation (SH) hotspot switching in L-shaped
plasmonic nanoantennas by means of spectral-phase shaping. L-shaped
nanoantennas were selected as simple model systems supporting multiple
plasmonic resonances within the spectrum of ultrafast Ti:Sapphire
lasers. Whereas polarization pulse shaping could be exploited as an additional
powerful degree of freedom, here we wanted to limit the complexity of the
optimization scheme.  Although sub-wavelength and second-harmonic hotspot
switching have been separately demonstrated~\cite{Aeschlimann2012,
  Brinks2013}, the control of SHG in sub-diffraction nanoantennas was not
shown before.

\section{Setup} \label{sec:setup}

In order to use NNs for optimal control on real nanoantennas, a reasonable
feasibility step is to train them using realistic simulated data. Producing
high-quality nanostructures and measuring them is still a time-consuming task,
but a vast simulated dataset can be produced in much shorter time. Plasmonic
nanostructures are particularly convenient to simulate: Their optical
properties are dominated by the surface density of charge and current, and it
is possible to accurately model them using a boundary element method (BEM)
\cite{Myroshnychenko2008}. For this purpose we applied a customized version of
the Matlab MNPBEM toolbox \cite{Hohenester2012}, which was extended to perform
non-linear optics simulations: additional details are provided in the appendix
in section \ref{sec:appendix:shg}.

\begin{figure}[htbp] \centering
\includegraphics[width=\linewidth]{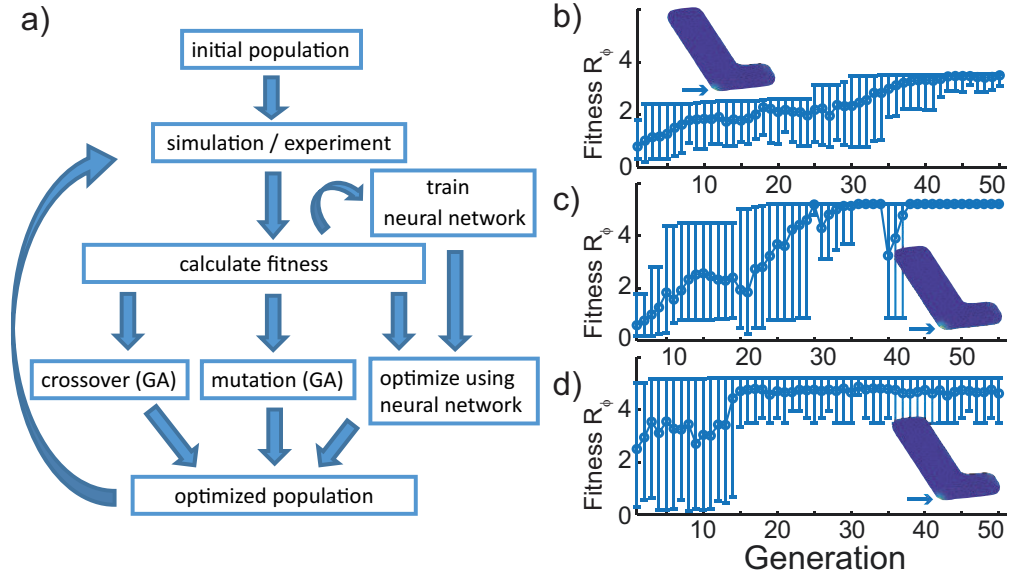}
\caption{a) Flow diagram of the training phase of the neural network using the
  population generated by a genetic algorithm. A random initial population of
  20 spectral phases is chosen. At each iteration, the fitness is calculated
  according to a pre-specified goal: The fittest individuals are kept
  unchanged and the rest undergo mutation or cross-over. The least fit
  individuals are optimized by the NN using back-propagation.  b) Training
  history of the GA without assistance from the NN. The graph reports the
  fitness of the best and worst individuals, and the average fitness. The
  fitness function (vertical axis) was chosen to maximize the relative SHG
  flux at a target hotspot of a $90~\mathrm{nm}\times250~\mathrm{nm}$ gold
  L-shaped nanoantenna, indicated by an arrow in the inset.  c) Similar to b)
  but with the GA assisted by an un-trained NN. d) similar to b) but with the
  GA assisted by a pre-trained NN: The fitness of the best individuals
  converges after just one iteration, the fitness of the worst individual
  remains lower due to the random mutations introduced by the GA.}
\label{fig:scheme_GA_NN}
\end{figure}

We trained the NN using the populations produced by a genetic algorithm, as
illustrated in Figure \ref{fig:scheme_GA_NN}a.  Using a GA has a double
advantage: First, it allows to have a fast feedback on the nanoantenna design,
e.g. if it is likely or not to be controllable withing the given experimental
parameters. Second, it generates a varied dataset of both near-optimal and
pseudo-random solutions.  At each iteration the population was used to train
the NN and then sub-divided according to the relative fitness: a few of the
fittest individuals were kept unchanged, 80\% of the remaining fittest were
used to create crossover children and the rest underwent random mutation. The
least fit individuals were optimized by the NN using a back-propagation
algorithm \cite{Lecun2015}. The choice of optimizing the least fit individuals
allows the possibility to start with an untrained NN.  As the NN was trained,
its accuracy increased until it started to accelerate the training of the
GA. When the NN was sufficiently trained, it could be directly used to produce
optimal solutions.

We note that an alternative and simpler way to generate a set of training
spectral phases would be to use a random numbers generator. However, random
phases tend to have fast oscillations that increase the pulse duration and
reduce any nonlinearity.

For the results in this paper, the NN was trained to predict the
second-harmonic (SH) flux at the hotspots on the surface of the nanoantenna
utilizing the spectral phase of the laser excitation pulse as input.  For
efficiency reasons, the spectral phase was specified at 6 nodal points, evenly
distributed within 2 standard deviations from the central frequency of the
laser excitation pulse. The phases were then interpolated on a finer mesh,
using piecewise cubic Hermite interpolation (pchip). An advantage of this
approach is that it produces reasonably smooth phase profiles, without fast
oscillations.  The spectral phase at the nodal points was bounded between
$\pm100~\mathrm{rad}$, an interval chosen to match the capability of a
standard $128~\mathrm{pixel}$ pulse-shaper. The spectral amplitude was taken
to be Gaussian with central frequency of $375~\mathrm{THz}$
($800~\mathrm{nm}$) and standard deviation of $26.5~\mathrm{THz}$,
corresponding to a full-width-half-max (FWHM) temporal intensity duration of
$10~\mathrm{fs}$. These constrains will make the results easier to test in an
experimental setup equipped with a pulse-shaper and a femtosecond laser
source.

\section{Results and Discussion}
The efficacy of the NN to accelerate the training of a GA is demonstrated in
Figure \ref{fig:scheme_GA_NN}. Panel b) shows the convergence of the GA
without assistance of the NN for a $90~\mathrm{nm} \times 250~\mathrm{nm}$
L-shaped gold nanoantenna. The bars indicate the fitness of the best and worst
individuals and the average fitness.  The fitness function was chosen to
maximize the relative SH flux at a specific hotspot, indicated in the inset by
an arrow: $R_{\phi} = h_{\phi, i} / \underset{i \neq j}{\max}(h_{\phi, j})$,
where $R_{\phi}$ is the fitness, $h_{\phi, i}$ is the SH flux at the target
hotspot '$i$' resulting from a specific spectral phase profile $\phi$ and
$\underset{i \neq j}{\max}(h_{\phi, j})$ is the maximum SH flux using the same
phase $\phi$ over all the other hotspots. The inset shows the generated
distribution of the SH field at the outer surface of the nanoantenna for a
flat phase pulse. Figure \ref{fig:scheme_GA_NN}c,d shows that a un-trained
neural network can reduce the number of needed iterations substantially and
that, with a pre-trained NN, the fitness of the best individual converges to
the optimal value in just one iteration.

Optimizing the relative local intensity of the SH allows to switch the
position of the brightest hotspot. 

\begin{figure}[htbp] \centering
\includegraphics[width=\linewidth]{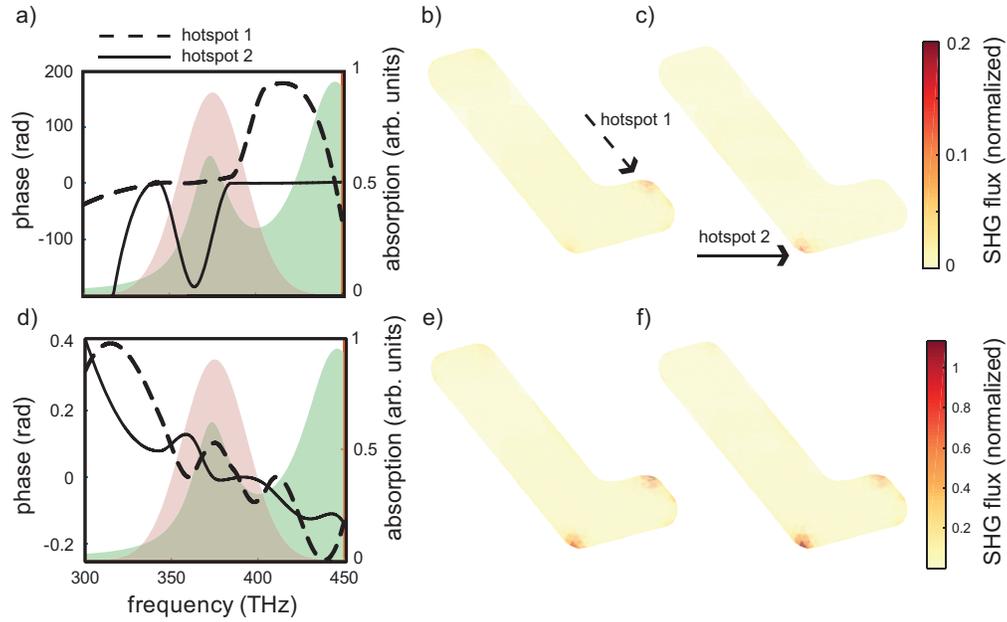}
\caption{Hotspot switching in a $90~\mathrm{nm} \times 250~\mathrm{nm}$ gold
  nanoantenna. a) Spectral phases which maximize the relative flux intensity
  of two hotspots (labeled '1' and '2') as found by the GA+NN algorithm. The
  hotspot positions are indicated by arrows in panels b,c).  The laser
  spectrum (orange shaded area) and the nanoantenna absorption spectrum (green
  shaded area) are also shown.  b,c) SHG flux intensity at the outer surface
  of the nano-antenna corresponding to the optimal phases shown in panel
  a). Panels d,e,f) report similar information as panels a,b,c) but with a
  different optimization goal: to maximize the absolute, rather than the
  relative, SH flux intensity at a specific hotspot }
\label{fig:hotspot_switching}
\end{figure}

The idea of hotspot switching is illustrated in Figure
\ref{fig:hotspot_switching}. Panel a) shows the laser spectrum, the
nanoantenna absorption spectrum and the spectral phases which maximize the
relative SH flux $R_{\phi}$ as defined above at the hotspots labeled '1' and
'2' in b) and c), respectively.  Panel b) shows the SH surface field: It can
be recognized that the location $x$ of the brightest hotspot changes according
to the optimization goal $F_{\phi}(x)$. The color scale was normalized to the
maximum SH surface flux for a flat phase laser pulse. It can be recognized
that maximizing the relative intensity of a specific hotspot comes with a
reduction of the overall SHG flux intensity of a factor of five.


Whilst optimizing the relative hotspot intensity is interesting, it
result in a overall decrease of the SH intensity, which might be detrimental
for actual applications. A different optimization goal is to maximize the
absolute value of the SH flux for a given hotspot.  The fitness is now defined
by $A_{\phi} = h_{\phi}/h_{0}$, where $h_{\phi}$ is again the SH flux at the
target hotspot resulting from a specific spectral phase $\phi$ and $h_{0}$ is
the SH flux over all the nanoantenna surface for a flat spectral phase
profile.  In this case we are not guaranteed that the target hotspot will be
the brightest one, but the resulting local fields will be larger.

We present the results of this kind of optimization in Figure
\ref{fig:hotspot_switching}d,e,f.  For the a 90 nm250 nm gold nanoantenna the
intensity gain was about a factor of 5 as compared to the optimization in
Figure 3b, c as can be seen from the respective color bars.

\begin{figure}[htbp] \centering
\includegraphics[width=\linewidth]{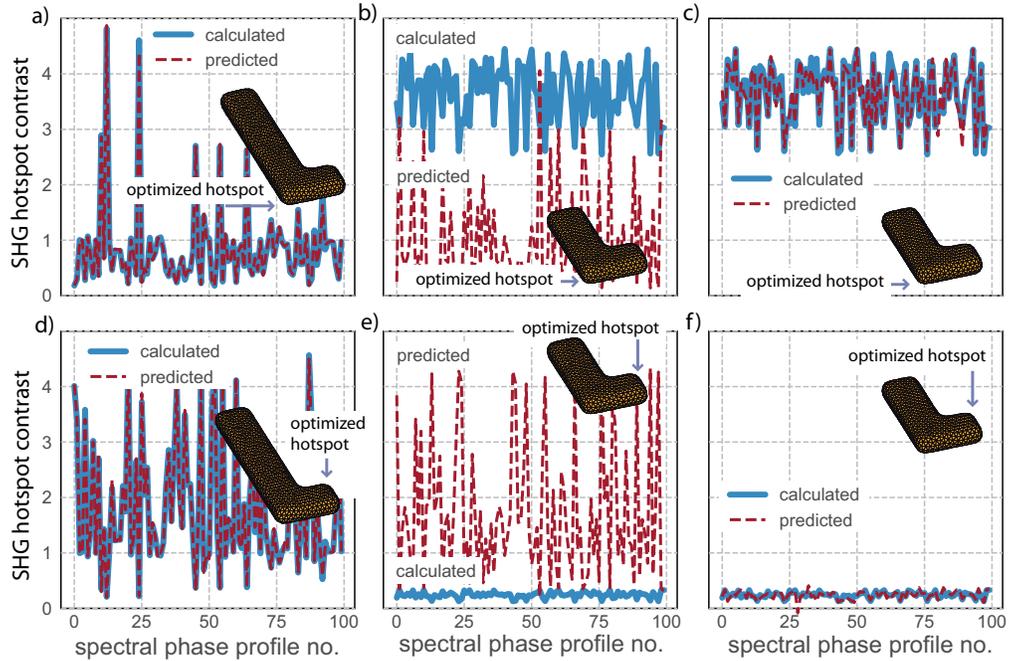}
\caption{a) Relative SH flux intensity for a target hotspot of a gold L-shaped
  nanoantenna with size $90~\mathrm{nm} \times 250~\mathrm{nm}$ for different
  spectral phase profiles of the incident laser pulse. The blue line refers to
  the simulated value, the red dashed line refers to the value predicted by
  the NN. The inset shows the nanoantenna with target hotspots marked.  Panel
  b,c) shows the performance of same NN but used to predict the SHG hotspot
  intensities for a nanoantenna with different size:
  $90~\mathrm{nm} \times 150~\mathrm{nm}$ with no further training, and after
  training only the output layer of the NN for 10 epochs. d,e,f) report
  similar results as a,b,c) but for a different hotspot. The figure
  illustrates the flexibility of the NN with respect to the sample size and
  aspect ratio.}
\label{fig:NNoptimization}
\end{figure}

The results shown in Figure \ref{fig:NNoptimization} demonstrate the
possibility of switching the position of the main hotspot between different
corners of an L-shaped nanoantennas. The displacement distance, for the
illustrated case, was $100~\mathrm{nm}$, well below the diffraction limit for
the laser excitation at $800~\mathrm{nm}$. The maximum contrast values,
integrated over the whole frequency range, were: $R_{\phi} = 5.2$ and
$R_{\phi} = 4.6$, for the two shown hotspots, which should be large enough to
be experimentally tested.  Such hotspot switching could be useful for
triggering local non-linear phenomena, either in a adjacent sample object,
such as a fluorescent molecule or semiconductor nanocrystal or, more simply,
in the substrate of the nanoantenna

Analyzing the spectral phase profiles found by NN-GA combination in
Figure~\ref{fig:hotspot_switching}a) and d) for the two different fitness
functions $R_{\phi}$ and $A_{\phi}$, we can comment on the mechanisms
underlying hotspot control in the present system.  In case of the optimization
of the relative flux $R_{\phi}$ at hotspot '1', the phase profile $\phi$
remains essentially flat between 320 and 390 THz, the spectral range of the
plasmon resonance connected to this particular hotspot. From 395 to 450 THz,
on the other hand, the phase becomes very large featuring huge higher order
variations. This leads to an efficient suppression of the SH flux of the
plasmon mode in this spectral range, which is associated with a different hotspot
distribution including a peak at position '2'. Optimizing the relative flux at
hotspot '2' is seen to result in the opposite spectral phase characteristics
with a flat curve between 450 to 395 THz and strong phase variations below.

For the target function $A_{\phi}$, which maximizes the absolute SH flux at a
given hotspot position, the observed phase variations are much smaller. Here,
the dominant control mechanism will be that of local pulse compression, also
mentioned in the introduction~\cite{Brinks2013,Huang2009, Brixner2006a}. In
other words, the optimum spectral phase profiles found by the NN+GA scheme
will compensate the local phase response of the nanoantenna at the particular
positions thereby maximizing the non-linear response.

An important advantage of NN-based control is expected to arise from the
adaptability of a trained network to similar problem sets.  Applied to
coherent control of plasmonic nanoantennas, this means that we could train our
NN on a specific nanoantenna and then use the learned weights also for
nanoantennas with different sizes or aspect ratios. The feasibility of this
idea is demonstrated in Figure \ref{fig:NNoptimization}. Panel a) shows the
prediction accuracy of a pre-trained NN for a set of randomly generated
spectral phases and for a nanoantenna of size
$90~\mathrm{nm} \times 250~\mathrm{nm}$, using a training set of $800.000$
phases, a test set of $90.000$ phases and $1000$ epochs.  The blue line refers
to the true value for the relative intensity of a target hotspot, marked with
an arrow in the inset of panel c). The red line shows the prediction by the
NN. The overall root square error was about $2.6\times10^{-2}$. Panel b) shows
the performance of the same NN but for a nanoantenna with size
$90~\mathrm{nm} \times 150~\mathrm{nm}$, and panel c) after fine tuning only the
last layer of the NN for $10$ epochs. The root mean square error was $0.98$
with no training and $5.7\times10^{-2}$ after fine tuning. The NN was still
able to quite accurately predict the relative intensity of the SH hotspot.
Panels d,e,f) report similar results, but for a different hotspot, which was
bright of the training antenna and dim on the test one: also in this case the
NN was able to make quite accurate predictions, with a relative error of about
$10$\%.

These results indicate that it should be feasible to use a NN trained with
simulated data to control a real nanoantenna in an experimental setup.  Due to
small changes in the dielectric environment and uncertainties in the
fabrication process, the plasmon frequencies could be different between the
real and the simulated nanoantennas. The ability of the NN to compensate for
the frequency shifts caused by changes in size and aspect ratio suggests that
the they could also adapt to shifts caused by the dielectric environment.

\section{Conclusion}
We introduced a novel scheme to control the near-field of plasmonic
nanostructures based on a neural network in conjunction with a genetic
algorithm. The neural network accelerates the optimization of the genetic
algorithm and stores information about the sample, which can readily be
generalized to other samples, with none or minimal further training.  In order
to prove the efficacy of this approach, we showed how the algorithms can find
the optimal spectral phases for switching the position of the brightest
hotspot in an L-shaped gold nanoantenna. We also showed that a NN trained on a
specific nanoantenna provides quite accurate results also for a nanoantenna
with different size and aspect ratio, even without any further training.  Our
results suggests that NNs are a powerful tool for optimal control of
near-fields at the nanoscale and, in perspective, for more complex
nanoplasmonics and nanophotonics devices.  Coherent control of hotspot
positions on the nanoscale could be experimentally observed using
photoemission electron microscopy (PEEM) or scanning near-field optical
microscopy with passive probes~\cite{Aeschlimann2012, Esteban2008}.  Possible
further developments include the coupling of nanoantennas to different
emitters or waveguides situated near the hotspots providing a means for
all-optical ultrafast switching or the spatially selective initiation of
photochemical reactions at plasmonic nanoantennas~\cite{Maier2018}.

\section{Appendix} \label{sec:appendix}

\subsection{Neural Network} \label{sec:appendix:NN} The neural-network (NN)
used to obtain the results shown in the article was a multi-layer perceptron
composed of four fully-connected layers, plus an input and an output
layer. The activation function was hyperbolic tangent (tanh) for all layers
except for the output layer, which had linear activation.  The input layer
contained 6 neurons representing the spectral phase of the excitation laser at
6 equidistant nodal points evenly distributed within 2 standard deviations
from the central frequency of the laser excitation pulse.  The phases were
then interpolated on a finer mesh, using piecewise cubic Hermite interpolation
(pchip).  The sizes of the inner layers were: 100, 100, 50, 50. The output
layer was used to predict the second-harmonic flux intensity for the 12
corners of a L-shaped nanoantenna.

The number of layers and neurons was chosen as a trade off between prediction
accuracy and training speed. A small network requires a relatively small
training dataset, and it can potentially be trained using experimental data
within a reasonable measurement time. The network was trained using a custom
implementation of the back-propagation algorithm in Matlab. The language was
chosen for easier integration with existing toolboxes for optics and
plasmonics. The accuracy of the code was tested using a widely-used
machine-learning toolbox (Keras2 with TensorFlow back-end).  The NN was
trained using using the population generated by a genetic algorithm
(GA). Several training sessions were run while changing optimization goals,
e.g. relative or absolute hotpot intensity for different choices of
hotspots. The final train set size was about 1 million, and the
test set size about 0.1 million. A small ($10^{-6}$) $L_{2}$ regularization
factor was used, however the regularization choice was not crucial when
training the NN using noiseless simulated data: It will be important for
training using real experimental data.

\subsection{SHG Simulations} \label{sec:appendix:shg} The second-harmonic
response of the L-shaped nanoantennas was calculated using the
boundary-element method, as described by Garcia de Abajo et
al. \cite{deAbajo2010}. An open-source Matlab toolbox (MNPBEM) was used to
calculate the linear response of the nanoantennas \cite{Hohenester2012}. The
surface density of charge and current was than used to estimate the dipolar
contributions to the SHG from each surface elements \cite{Bachelier2010,
  Czaplicki2015}. For simplicity, the bulk contribution to the SHG was not
considered. The simulation was performed in frequency domain using 200 points
between $200~nm$ and $1000~nm$, the results were then interpolated using a
finer mesh over the spectral range of the laser excitation.  The simulated
laser pulse was a Gaussian with temporal full-width-half-maximum width of
$10~\mathrm{fs}$ and central frequency of $375~\mathrm{THz}$
($800~\mathrm{nm}$).

\section{Funding Information}

Deutsche Forschungsgemeinschaft (DFG) HA4405/8-1

\section{Acknowledgments} The authors thanks Giovanni Piredda, Nicolas Coca
Lopez, Veit Giegold and Richard Ciesielski for helpful discussions.

\end{document}